\documentclass[epj]{webofc}
\usepackage[varg]{txfonts}   
\woctitle{International Conference on New Frontiers in Physics 2013}
\newcommand{\mt}{$m_{\mathrm{T}}$}
\newcommand{\kt}{$k_{\mathrm{T}}$}
\newcommand{\apap}{$\bar{\mathrm p}\bar{\mathrm p}$}
\newcommand{\pap}{p$\bar{\mathrm p}$}
\newcommand{\pala}{p$\bar{\mathrm \Lambda}$}
\newcommand{\laala}{$\Lambda\bar{\mathrm \Lambda}$}
\newcommand{\apla}{$\bar{\mathrm p} \Lambda$}
\newcommand{\apala}{$\bar{\mathrm p} \bar{\Lambda}$}
\newcommand{\snn}{$\sqrt{s_{\mathrm {NN}}}$}

\begin{document}
\title{Baryon femtoscopy in heavy-ion collisions at ALICE}
%
%

\author{Maciej Pawe$\l$ Szyma\'nski \inst{1}\fnsep\thanks{\email{maszyman@if.pw.edu.pl}} for the ALICE Collaboration
}

\institute{Faculty of Physics, Warsaw University of Technology, Koszykowa 75, 00-662 Warsaw, Poland}

\abstract{%
In this report, femtoscopic measurements with pp, \apap, \pap, \pala, \apla~and \laala~pairs in {Pb--Pb} collisions at \snn$=2.76$~TeV registered by ALICE at the LHC are presented. Emission source sizes extracted from the correlation analysis with (anti)protons grow with the event multiplicity, as expected. 
A 
method to extract the interaction potentials (e.g. for \pala~and \apla~pairs) based on femtoscopy analysis is discussed. The importance of taking into account the so-called residual correlations induced by pairs contaminated by secondary particles is emphasized for all analyses mentioned above.
}
\maketitle
\section{Introduction}
\label{intro}
ALICE, A Large Ion Collider Experiment at the Large Hadron Collider (LHC) at CERN is dedicated to the study of the properties of the quark-gluon plasma (QGP), the state of matter described by partonic degrees of freedom~\cite{alice}. Such a state may be created in ultra-relativistic collisions of heavy ions. The information about the spatio-temporal characteristics of the final hadronic system of the collision can be deduced from analysis techniques~\cite{kopylov2} based on the measured two-particle correlations at low relative momenta and the knowledge of their sources; namely, Final State Interactions (Coulomb and Strong) and Quantum Statistics (in the case of identical particles). 
 The two-particle correlation can be expressed by the following equation~\cite{lednicky_finite_size}:
\begin{equation}
  C({k^*}) = \int S({k^*},{r^*}) \Psi ({k^*},{r^*}) \mathrm{d}^4{r^*},
\label{eq:koonin}
\end{equation}
where $C$ is the measured correlation, ${k^*}$ is half of the pair relative momentum, ${r^*}$ is the pair relative separation, $S$ is the source function and $\Psi$ is the pair wave function. 
Generally, the two-particle interaction is known for the most commonly analysed types of pairs, e.g. for pions and kaons. Hence, the measured correlation function may be used to extract the information about the source function (e.g. a Gaussian shape of the source can be assumed to determine the width of the distribution). For instance, $\Psi$ for pp pairs contains components from Fermi-Dirac statistics, Coulomb interactions (both leading to anticorrelation) as well as strong Final State Interactions (resulting in a characteristic peak at ${k^* \approx 20}$~MeV/$c$, whose height is sensitive to the femtoscopic radius).

The goal of the baryon-(anti)baryon femtoscopic analysis is to complement the measurements of the transverse mass (\mt) dependence of the ''homogeneity lengths''~\cite{sinyukov} (the sizes of the phase space of emitted particles with specific velocities) extracted so far from correlations of pions and kaons. Based on such measurements, one is able to verify the \mt~scaling of the source size~\cite{lisa}, commonly explained in the framework of hydrodynamic models as a signature of collective behaviour of matter created in heavy-ion collisions.

However, the strong FSI parameters for many systems such as \pala, \apla~and \laala~are not known precisely. Experimental data of 
cross-sections for these pairs are very limited. Only phenomenological models exist, predicting the values of the parameters of the strong interaction~\cite{bbint1, bbint2, bbint3, bbint4}. Therefore, femtoscopic techniques may be helpful in constraining the strong FSI parameters for the aforementioned systems. This can be achieved by inverting the standard femtoscopic procedure by fixing $S$ in Eq.~\eqref{eq:koonin} to extract $\Psi$. For instance, one can measure the \apla~correlations and assume that the source size for this pair is the same as for the pp system (it should be similar indeed, due to the comparable masses of the p and $\Lambda$ baryons). Then, the Lednicky-Lyuboshitz analytical model~\cite{lednicky} might be used to infer the information about the FSI parameters. In this model the correlation function is calculated as the square of the wave function averaged over the total spin and the relative separation of the points of particles emision in the pair rest frame. One should emphasize that the results of such an analysis can be applied in modelling the phase of hadronic rescatterings. In particular, one can use the extracted values of the parameters which account for annihilation in the baryon-antibaryon systems. 
This may explain 
the observed decrease of baryon yields which are below expectations from thermal models at the LHC energies~\cite{alicespectra, karpenko}. Also, more precise knowledge of the parameters of strong interactions for baryon pairs is of great importance for astrophysics; in particular in understanding the properties of neutron stars~\cite{neutron}.


The aforementioned analyses require taking into account the so-called residual correlations~\cite{wangrescorr, stavrescorr, stavrescorr2}. The reason for that is the following: in  heavy-ion collisions in a collider setup, a significant number of secondary baryons (products of weak decays) cannot be distinguished from the primary ones (prompt particles produced in the collision). Therefore, a sizeable number of analysed pairs are composed of at least one secondary particle. 
Hence, the correlations of such pairs arise due to the interactions between the parent particles. The effect becomes more and more important when the momentum of the decay products, in the rest frame of the parent particle, is relatively small. Then, the daughter particle will move in a direction similar to the direction of the parent particle and the correlation, which diluted somewhat, will still be observed for such a pair.

\section{Data analysis}
\label{sec-1}
The analysis was performed over a sample of about 30 million {Pb--Pb} collisions at {\snn$=2{.}76$~TeV} registered by ALICE~\cite{alice}. The VZERO detector (the forward scintillator arrays) was used for centrality determination and triggering. Events with an interaction vertex within 8~cm from the nominal interaction point in the beam axis were selected. The Time Projection Chamber (TPC) was used for track reconstruction. The identification of (anti)protons was based on the measurements of the specific ionisation energy loss by the TPC and the time-of-flight by the TOF and T0 detectors, in conjunction with the momentum of the particle inferred from the curvature of its trajectory in the magnetic field. 
$\Lambda$ and $\bar{\Lambda}$ baryons were selected using their decay topology by identifying the daughter tracks measured in the TPC and TOF. Only particles within the pseudorapidity range $|\eta| < 0.8$ were accepted in the analysis which corresponds to the region of uniform TPC acceptance. To reduce the contribution from secondary particles, a cut on the distance of closest approach of the particle trajectory to the primary vertex was applied.

The correlation function was obtained by dividing the signal and the uncorrelated distributions. The former was formed by calculating the relative momentum (in one dimension, in the Pair Rest Frame) $q_{\mathrm{inv}}=2 \cdot k^{*} = |q-P(m_1^2-m_2^2)/P^2|$ ($q$ is the pair relative momentum, $P$ is the pair total momentum and $m_1$, $m_2$~denote the masses) for particles from the same event. The uncorrelated distributions were created by pairing particles from different events. Pair selections were applied to account for fake pairs with low relative momentum due to the so-called splitting (i.e., one particle reconstructed as two tracks) and two-track inefficiency due to merging, (i.e., two tracks reconstructed as one). These selection criteria made use of the ratio of the detector signals (clusters) shared by two tracks to all clusters and the angular distance between two tracks inside the TPC.



\section{Results}
\label{sec-2}
\begin{figure}[h]
  \centering
  \includegraphics[width=0.49\textwidth]{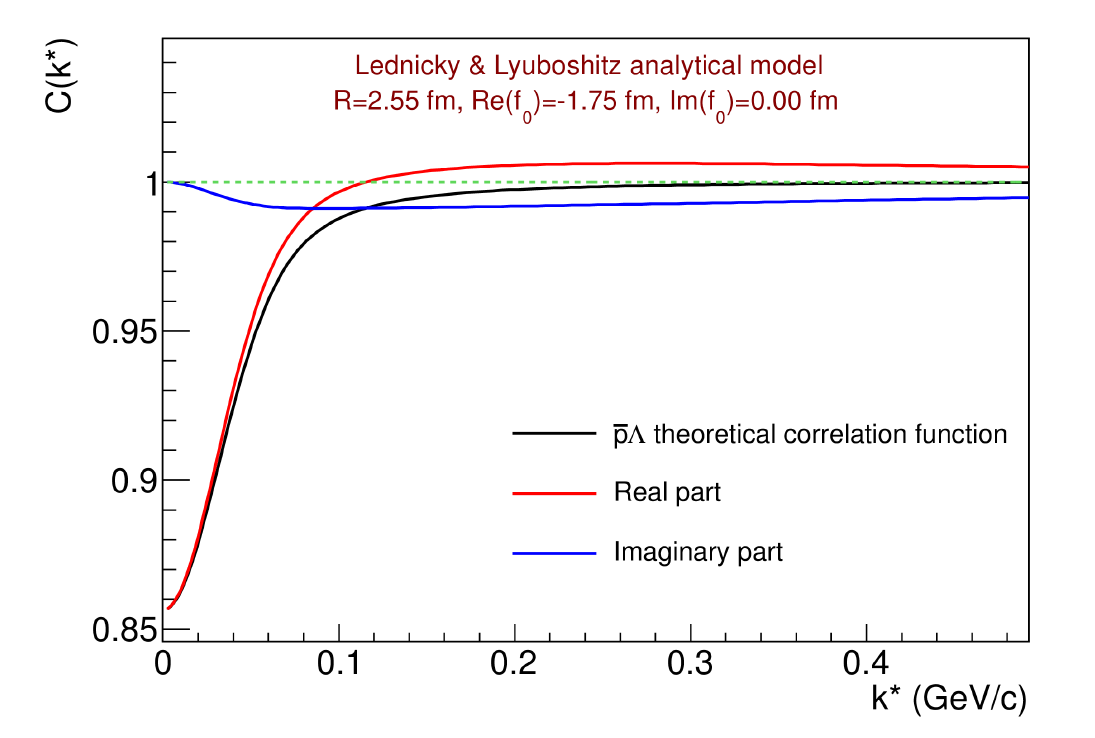}
  \includegraphics[width=0.49\textwidth]{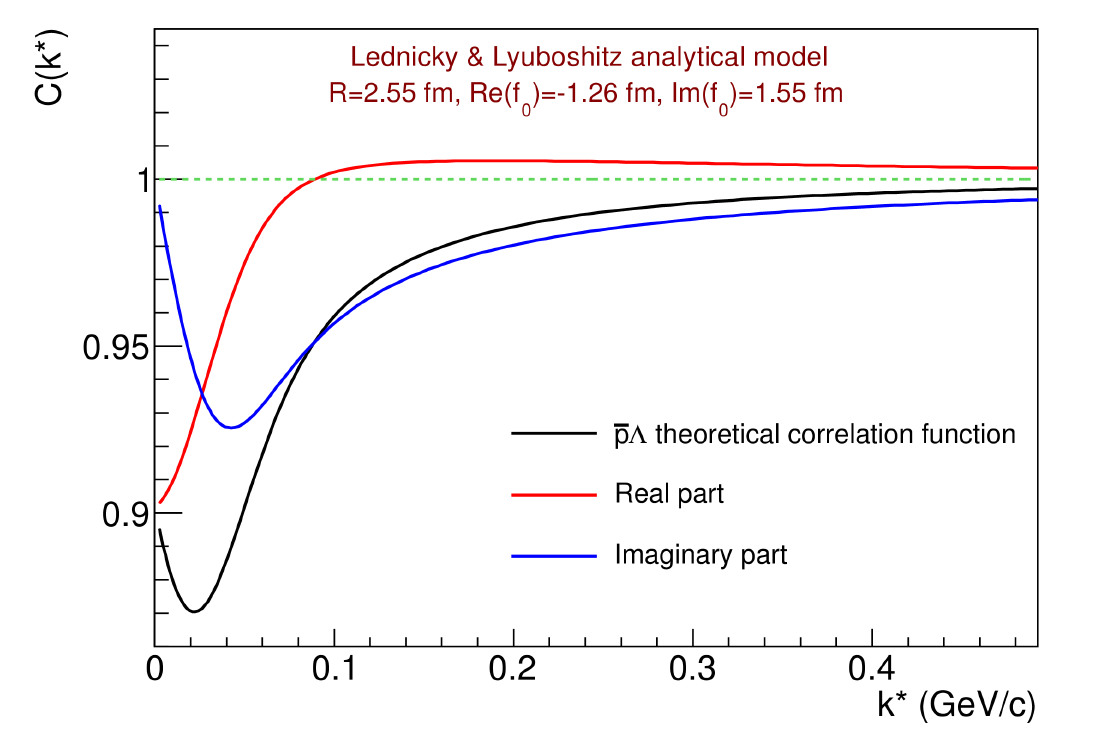}
  \caption{\apla~correlation functions and their components calculated from the Lednicky-Lyuboshitz analytical model for different FSI parameters (see text for details). The left plot shows the results of calculations with the imaginary part of $f^S(k^{*})$ set to zero. The right plot is the example of the calculations with the non-zero imaginary part of $f^S(k^{*})$.  }
  \label{llapla}
\end{figure}

In fig.~\ref{llapla}, \apla~correlations obtained using the Lednicky-Lyuboshitz analytical model~\cite{lednicky} are shown. They are calculated using the formula given by Eq.~\eqref{eq:koonin2}.
\begin{equation}
  C({k^*}) = 1+ \sum_{S} \rho_S  \left [{1 \over 2} \left| {f^S(k^{*})}\over{R}\right|^2  \left( 1-{{d_0^S}\over{2 \sqrt{\pi} R} } \right)  + {{2 \Re f^S(k^{*})}\over{\sqrt{\pi} R}} F_1 (2k^{*}R) - {{\Im f^S(k^{*})}\over{R}}F_2(2k^{*}R) \right ],
\label{eq:koonin2}
\end{equation}
where $F_1(z) = {\int_0^z \mathrm{d}x {{\mathrm{e}^{x^2-z^2}}\over{z}}}$, $F_2(z)=1-{{1-\mathrm{e}^{-z^2}}\over{z}}$, $f^S(k^{*}) = (1/{f_0^S+{1 \over 2} d_0^S k^{*} -ik^{*}})^{-1}$ is the spin-dependent scattering amplitude, $\rho_s$ is the fraction of pairs in each total spin state S. In this analysis only spin-averaged values are considered. The effective radius of the interaction $d_0^S$ is set to zero, following the procedure applied in~\cite{starpla}. Therefore, the scattering amplitude depends on the scattering length $f_0^S$. The contribution from the real part of the scattering amplitude can lead to either positive or negative correlation but it is always narrow in $k^{*}$ 
in comparison to the imaginary part
, which accounts for annihilation in the Final State Interactions. Therefore, only the non-zero $\Im f^S(k^{*})$ may lead to an anticorrelation wide in $k^{*}$.

\begin{figure}[h]
  \centering

  \includegraphics[width=0.99\textwidth]{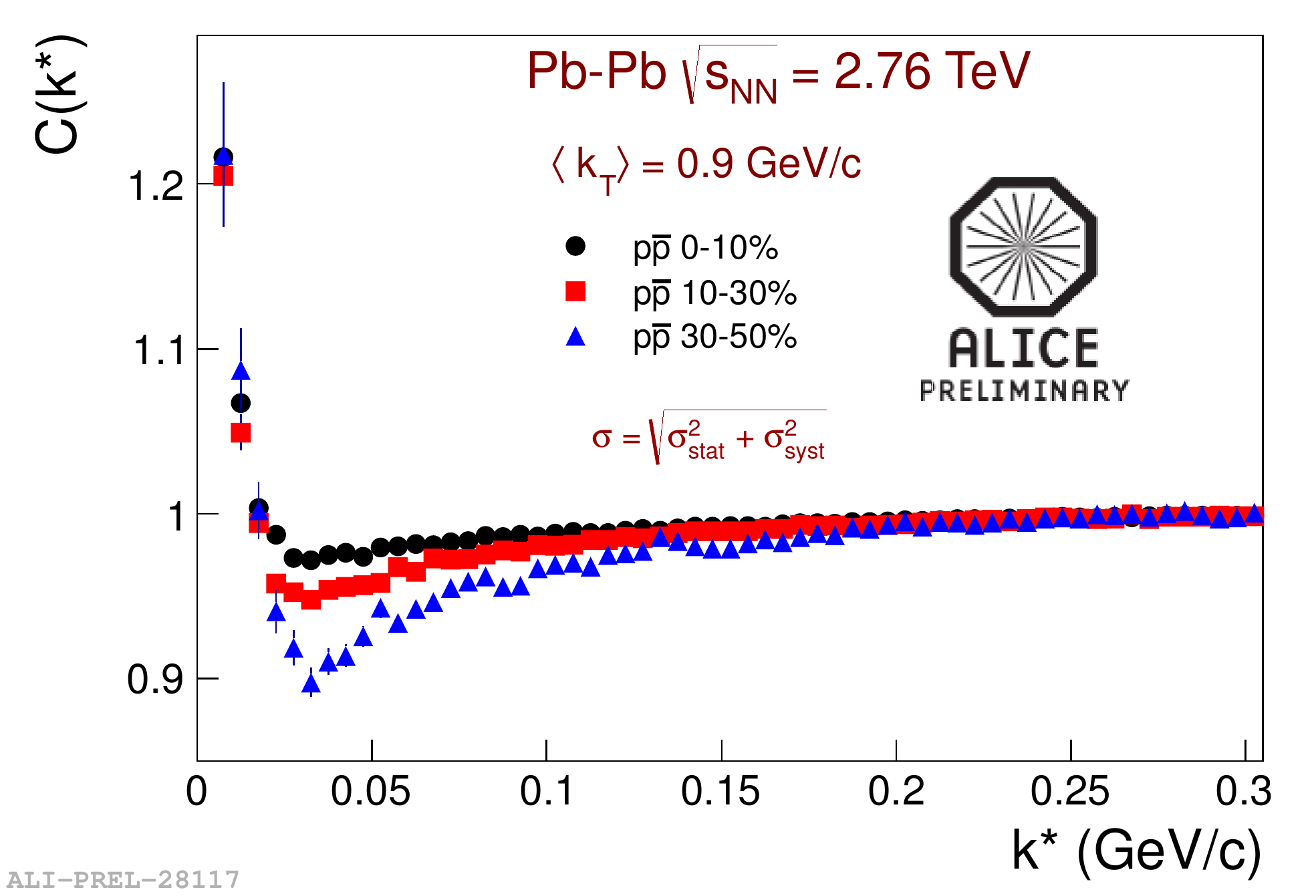}

  \caption{\pap~~correlation functions from the Pb--Pb collisions at \snn$=2.76$~TeV.}
  \label{cfpap}
\end{figure}

The \pap~correlation functions, for different centrality classes are presented in fig.~\ref{cfpap}. As extensively studied and discussed in the literature~\cite{lednicky_finite_size}, they show a peak at low relative momenta, due to the Coulomb attraction. For larger values of $k^{*}$ a wide anticorrelation is observed. It is caused by the annihilation component of the strong FSI, similarly to the \apla~case described above with the Lednicky-Lyuboshitz analytical model. Both effects are necessary to describe the data which is compatible with the baryon annihilation in the final state. Moreover, one can notice that the anticorrelation is stronger for more peripheral events which reflects the smaller size of the source.



The \apla~and \pala~correlation functions are shown in  fig.~\ref{cfapla
}. For these systems, the strong FSI is the only source of femtoscopic correlations. No significant difference between \apla~and \pala~ is observed, as expected. A wide negative correlation can be noticed. Based on the explanation by the Lednicky-Lyuboshitz model, it is qualitatively consistent with the annihilation from the FSI.
\begin{figure}[h]
  \centering
  \includegraphics[width=0.99\textwidth]{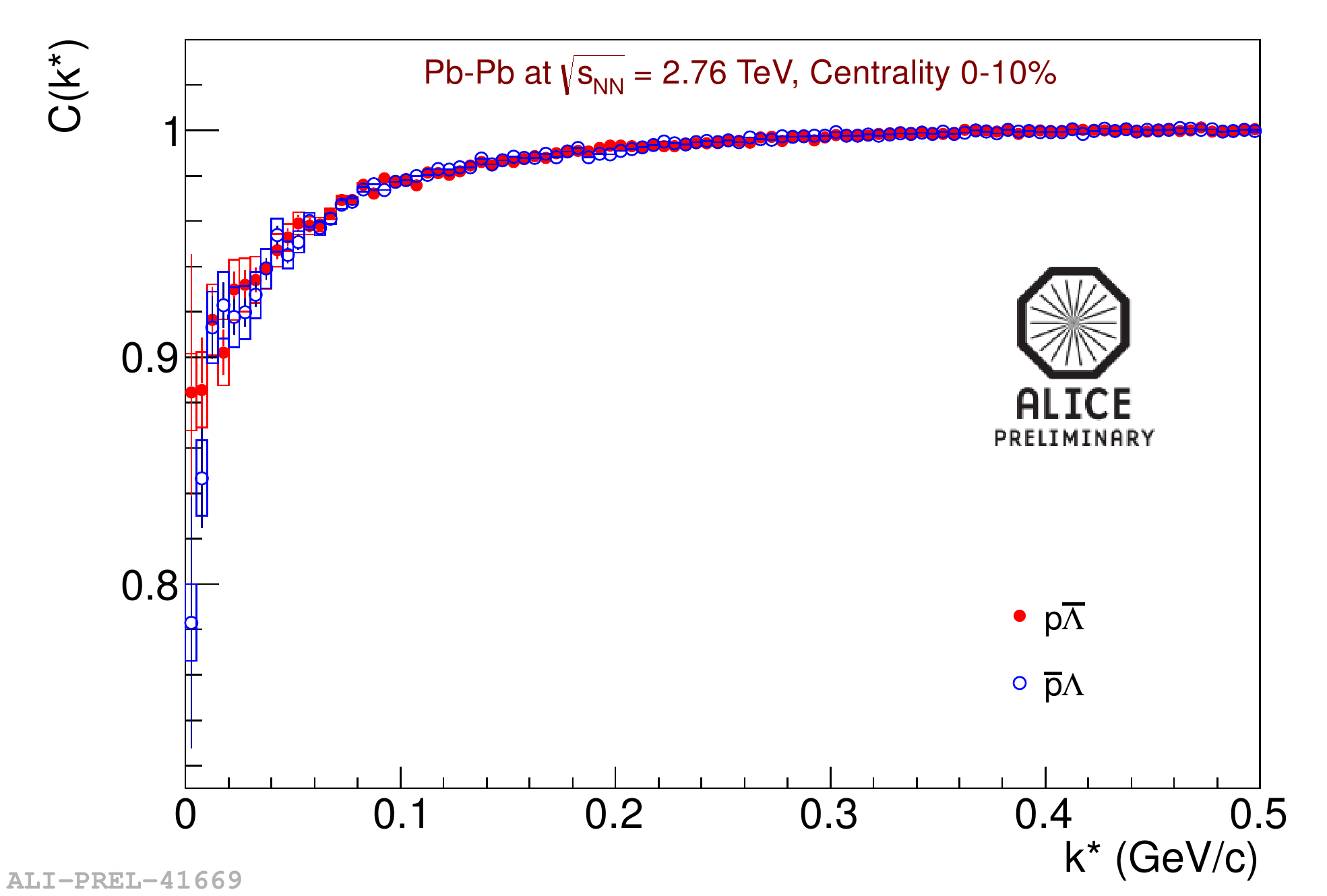}
  \caption{\pala~and \apla~correlation functions from the Pb--Pb collisions at \snn$=2.76$~TeV.}
  \label{cfapla}
\end{figure}

In fig.~\ref{cflaala
}, the \laala~correlation functions are presented. The suppression at low relative momenta for three centrality ranges is clearly visible. Again, it is compatible with baryon-antibaryon annihilation (see above). The strength of the correlation decreases for more central events which indicates that the emitting source size becomes larger for such events.
\begin{figure}[h]
  \centering
  \includegraphics[width=0.99\textwidth]{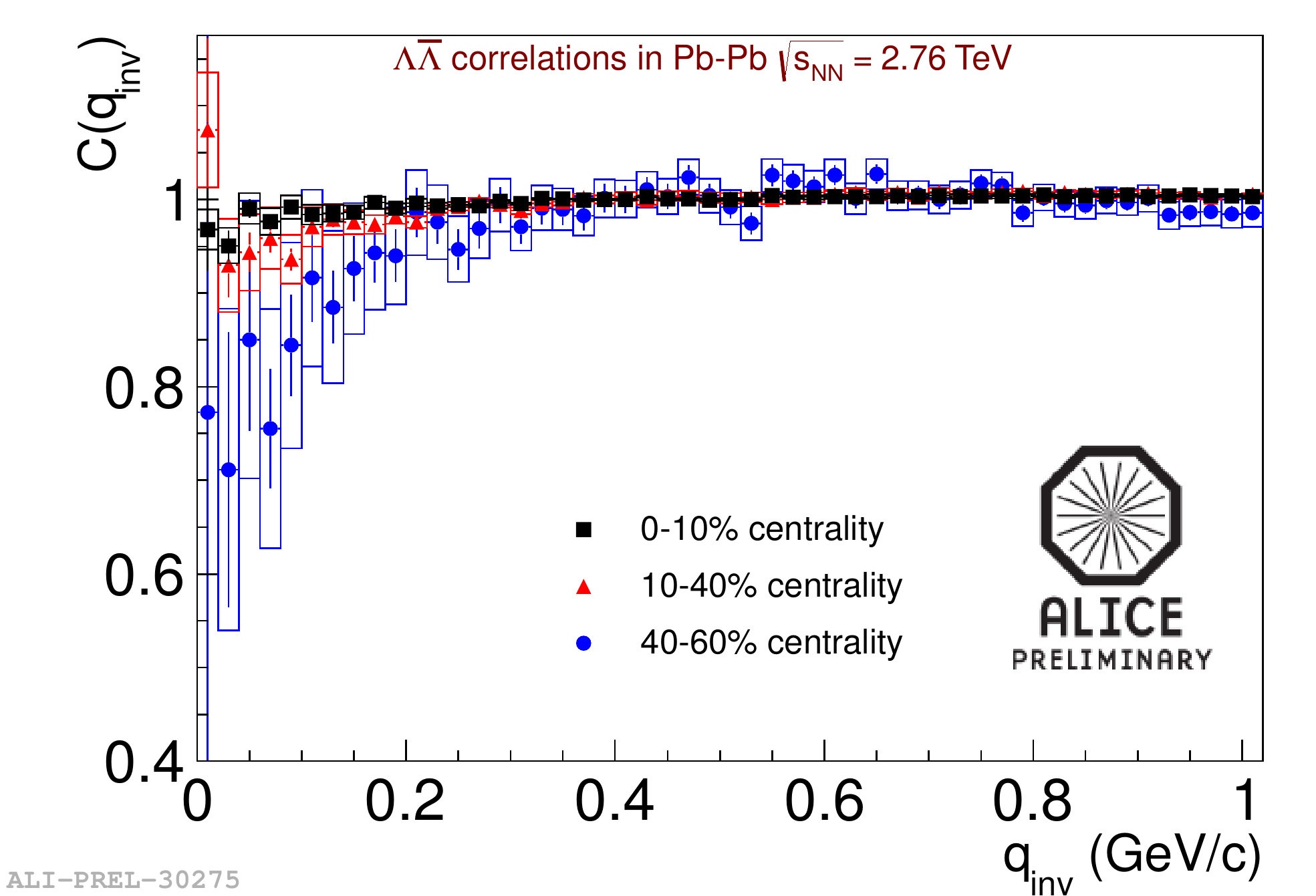}
  \caption{ \laala~correlation functions from the Pb--Pb collisions at \snn$=2.76$~TeV.}
  \label{cflaala}
\end{figure}

\begin{figure}[h]
  \centering
  \includegraphics[width=0.49\textwidth]{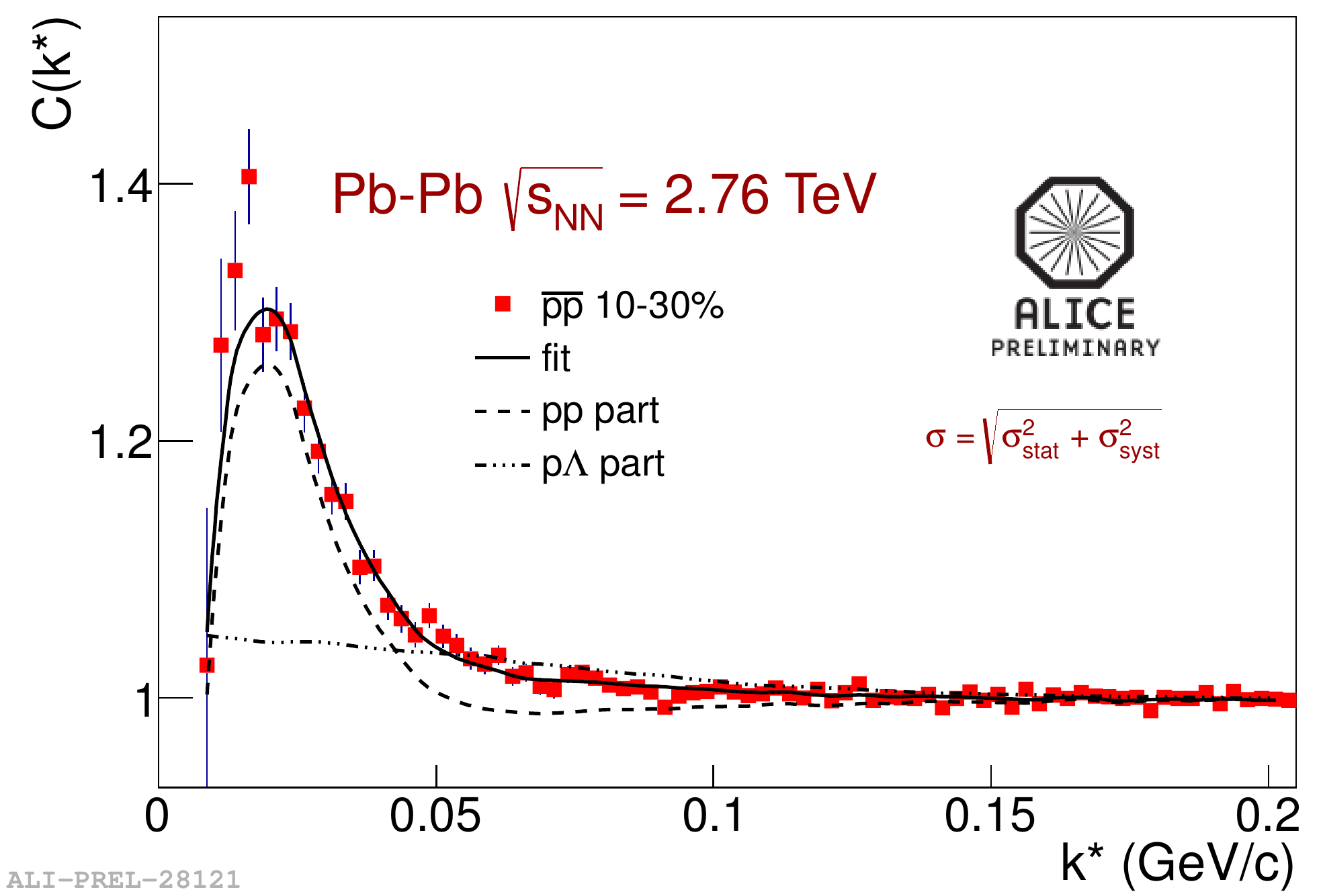}
  \includegraphics[width=0.49\textwidth]{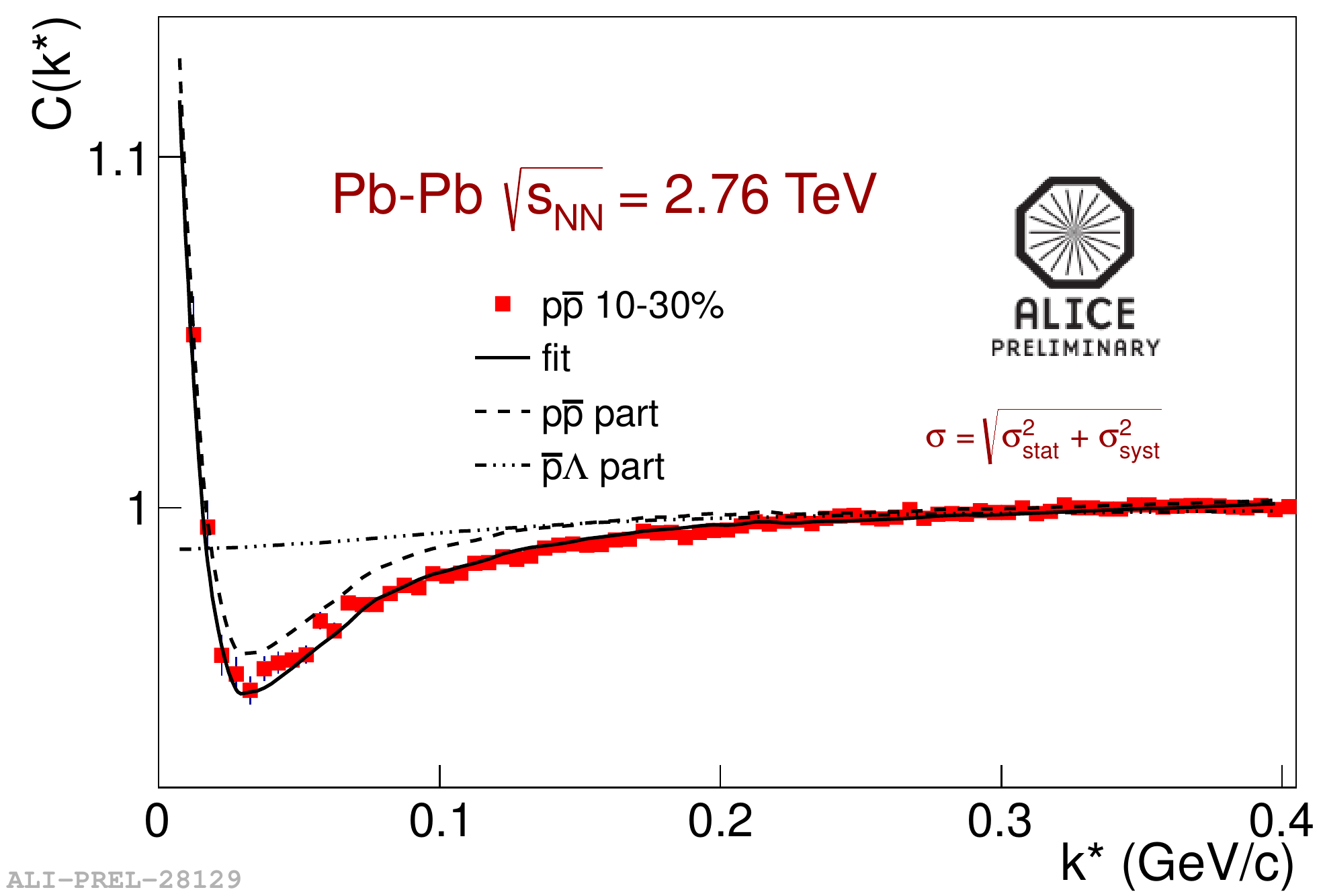}
  \caption{Example of the fit to the \apap~and \pap~correlation function, taking into account contributions from residual correlations.}
  \label{apapfit}
\end{figure}

The femtoscopic correlations for pp and \apap~(shown in the left panel of fig.~\ref{apapfit}) are due to the interplay of Fermi-Dirac statistics, Coulomb and Strong FSI resulting in a distinctive maximum for ${q_{\mathrm {inv}}=2 k^{*}~\approx~40}$~MeV/$c$ \cite{lednicky}. Since the feed-down from weak decays cannot be neglected in ultra-relativistic heavy-ion collisions, the residual correlations from the p$\Lambda$ system in the pp correlations ought to be taken into consideration. As a $\Lambda$ baryon decays into a proton and a $\pi^{-}$ with a small decay momentum with respect to the proton mass, femtoscopic correlations between a primary p and a $\Lambda$ may be detected for a pair composed of the primary p and the secondary p from the $\Lambda$ decay. The method of simultaneously fitting the pp (\apap) and p$\Lambda$ (\apala) correlations was applied to extract the femtoscopic radii from proton femtoscopy. As one can observe in fig.~\ref{apapfit} the contribution from genuine \apap~correlations in the measured correlation function describes the maximum at $k^{*}~\approx~20$~MeV/$c$. Also the wide correlation excess is fairly well reproduced by the residual correlations from the \apala~system. The same combination of effects was needed to describe the \pap~correlations (see the right panel of fig.~\ref{apapfit}).

The shape and strength of these residual correlations might be sensitive to the FSI parameters, though. One can incorporate these parameters from existing models and particles' momenta generated by a model of heavy-ion collision (e.g. \verb|THERMINATOR 2|~\cite{therminator2}) to simulate the impact of residual correlations on the measured correlations for baryon pairs. Then, the same analysis procedure (as the one depicted in fig.~\ref{apapfit}) may be performed for the obtained correlation functions to fit not only the radii but also the FSI parameters. Therefore, the described method may be used to constrain their values. This is especially important for baryon-antibaryon pairs where the FSI parameters are known with large uncertainty or even unknown.



\begin{figure}[h]
  \centering
  \includegraphics[width=0.99\textwidth]{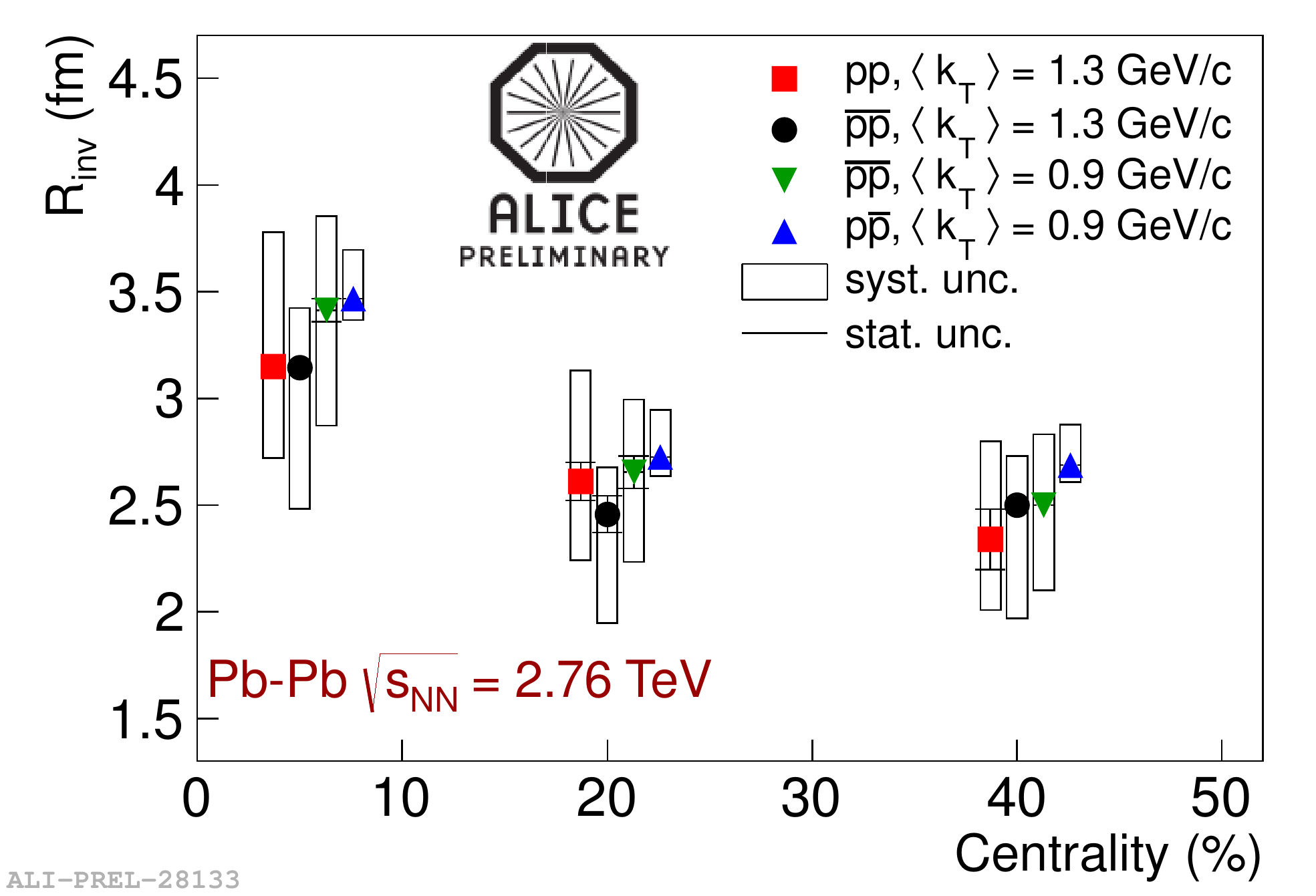}
  \caption{Pair transverse momentum dependence of the radius parameter extracted from correlations of protons.}
  \label{ktdep}
\end{figure}

The pair transverse momentum (\kt=${1 \over 2} {|\vec{p}_{\mathrm{T,1}}+\vec{p}_{\mathrm{T,2}}}|$)~dependence of the invariant radii deduced from proton femtoscopy is presented in fig.~\ref{ktdep}. The extracted radii increase with event multiplicity and slightly decrease with \kt.

\section{Summary}
\label{sec-3}
Femtoscopic correlation functions for baryon-(anti)baryon pairs are presented. The evident anticorrelation observed in \pap, \pala, \apla~and \laala~correlations is consistent with the hypothesis that annihilation from the strong FSI may cause that lower baryon yields are observed compared to thermal models at LHC. The importance of the residual correlations in baryon femtoscopy is stressed. The radii deduced from proton femtoscopy analysis show an increase with event multiplicity and decrease with pair transverse momentum.

\section*{Acknowledgements}

This work has been financed by the Polish National Science Centre under decision no. 2011/01/B/ST2/03483. 

%
%
%

\end{document}